\documentclass[12pt,a4paper,english]{article}
\usepackage[T1]{fontenc}
\usepackage[latin9]{inputenc}
\usepackage{textcomp}
\usepackage{amsmath}
\usepackage{graphicx}
\usepackage{setspace}
\usepackage{amssymb}

\makeatletter

\newcommand{\theAddress}[1]{
\par {\raggedright #1
\vspace{1.4em}
\noindent\par}
}

\usepackage{babel}
\makeatother

\begin{document}

\title{An Analytical Study in Coupled Map Lattices of Syncronized States
and Travelling Waves, and of their Period-Doubling Cascades}

\author{Mª Dolores Sotelo Herrera${}^{a}$ \& Jesús San Martín${}^{a,b}$}

\date{~}

\maketitle

\theAddress{${}^{a}$ Departamento de Matemática Aplicada, E.U.I.T.I., Universidad
Politécnica de Madrid. Ronda de Valencia 3, 28012 Madrid Spain\\
${}^{b}$ Departamento de Física Matemática y de Fluidos, U.N.E.D.
Senda del Rey 9, 28040 Madrid Spain\\
Corresponding author: jsm@dfmf.uned.es}

\begin{abstract}
Several theorems are demonstrated that determine the sufficient conditions
for the existence of synchronized states (periodical and chaotic)
and also of travelling waves in a CML. Also are analytically proven
the existence of period-doubling cascades for the mentioned patterns.
The temporal state of any oscillators are completely characterized.
The given results are valid for a number of arbitrary oscillators
whose individual dynamics is ruled by an arbitrary $C^{2}$ function.
\end{abstract}
Systems showing patterns as a consequence of the interaction among
their diverse components are really frequent, in any field that one
can imagine: neuronal activity within the brain, or the function of
organs as a whole within the body, drivers on a motorway, birds flying
in a group, a network of computers, coupled lasers, crystal growth,
etc. 

The result of the interaction of the individual elements generates
structures that manifest in the system as a whole. In these processes,
one should consider two things: the behavior of any individual and
the interaction among them. If we consider the traffic example, it
is clear that the behavior of an individual driver, that is his decision
to drive in a particular way or another, is certainly different when
there are few cars on a motorway or when there is a traffic jam (in
which case he will be guided by traffic patterns). 

Broadly speaking, all of these systems consist of a group of elements
coupled by some kind of process, and at the same time, every element
of the group is ruled by its own local dynamics. The understanding
of such systems is extraordinarily complicated, since there are no
particular mathematical tools developed to study them. One way to
confront this problem is to discretize spatial and temporal variables
as well as to fix inter-individual interactions as well as the individual
dynamics. The result is a Coupled Map Lattice (CML) \cite{libro kaneko}:
a chain of coupled elements (called oscillators), each situated on
a discrete point of the lattice, whose individual dynamics is ruled
by a discrete map. Despite the spatial and temporal variables are
discretized, state variables remain continuous.

In the last few years, CML have been extensively studied since the
work of Kaneko and colaborators \cite{Kaneko89,Kaneko90a,Kaneko90b,kaneko91a,kaneko91b},
and from the beginning, they have shown themselves to be exceptional
modelling spatially extended systems. The use of this study has been
extended into diverse scientific branches with an extraordinary variety
of applications in physics, biology, chemistry, social sciences, and
engineering modeling. \cite{PhysicaD,Chaos}

~

A typical evolution equation for a CML \cite{libro kaneko} is given
by

~\begin{equation}
X_{i}(n+1)=(1-\alpha)f(X_{i}(n))+\frac{\alpha}{m}\sum_{j=1}^{m}f(X_{j}(n))\label{eq:uno}\end{equation}

\[
i=1,...,m\]
where $X_{i}(n)$ represents the state of the oscilator located at
node {}``i'' of the lattice, in the instant {}``n''. The parameter
$\alpha$ weights the coupling among oscilators. Periodic conditions
are assumed in the boundaries, given as \[
X_{i}(n)=X_{i+m}(n)\;\;\forall i\]
Depending on the value of $\alpha$ , the system behavior changes
from the independent evolution of each oscilator (for $\alpha=0$)
up to a mean field approach (for $\alpha=1$). For intermediate values
$0<\alpha<1$ the system is ruled by both local and global mechanisms.

The general form of the coupling term is given by\[
\frac{\alpha}{m}\sum_{j=1}^{m}w_{ij}f(X_{j}(n))\]
where the $w_{ij}$ measure the weights between the $j$-th oscilator
and the $i$-th one. To achieve a symmetrical and spatially invariant
coupling, it is usually taken $w_{ij}=\bar{w}_{\vert i-j\vert}$.
Sometimes, the coupling term will be written as \[
\frac{\alpha}{m}\sum_{j=1}^{m}f(X_{j}(n))\]
(mean field), or \[
\frac{1}{2}\left[f(X_{j-1}(n))+f(X_{j+1}(n))\right]\]
(nearest-neighbor coupling). However, this last description is not
adequate when we are dealing with a supercritical bifurcation threshold,
because the coherence lengths are usually quite large \cite{Chate1988}.
Given that, in this paper, we want to study bifurcations in CML, we
will use the mean field approach.

Another important point, that must be considered, is the updating
of oscillators; they can be synchronous (all oscillators are updated
simultaneously) or asynchronous (oscillators are updated one at a
time) \cite{Atmans1,Mehta}. Choosing one or the other depends on
whether oscillators communicate among them much quicker that the updating
time of the system as a whole, which is ruled by the evolution equation
\eqref{eq:uno}. In this paper we will refer to synchronous systems.

In the scientific literature, the majority of the results, referring
to CML, are numerical results, as we will see later. The awesome richness
of numerical results is restricted by a fixed and finite set of parameter
values, and a finite number of oscillators in the CML, which supposes
a limitation for adequate understanding of certain phenomena. In particular,
the transition to chaos by period duplication needs the period to
tend to infinity. It is also necessary the number of oscillators to
be infinite, in a finite region, for the understanding of the onset
of turbulence in fluids and plasmas; otherwise, there would be a cutoff
in the wave numbers that could be studied because the lattice would
have a finite spatial resolution. Mathematical proofs would be desirable
to characterize syncronized states, traveller wave bifurcations and
other behaviours. Fortunately, numerical results point out us what
to look for and where.

In this paper, analytical proofs, in CML, of the existence of syncronized
states and travelling waves will be given. It will be proved that
both patterns will go under a period doubling cascade as $f$ , in
\eqref{eq:uno}, does. These behaviours will be completely characterized,
giving analytical expressions of the temporal evolution of every oscillator.

The fixed points of CML, generated in period-doubling cascades, will
be essentially the fixed points of $f^{m2^{k}}$($m$ number of oscillators
in CML). As $f$ determines the individual dynamics (see (\ref{eq:uno})),
what is shown is the emergence of global properties from the local
dynamics of a single oscilator.

We have tried to keep the widest generality in the results; therefore,
theorems have been proved using an arbtirary $C^{2}$ function $f(x;r)$,
instead of working with the logistic equation (or any toplogically
conjugated functions) as usual.

Perturbative methods will be used to obtain analytical solutions.
The inversion of functional matrices of arbitrary size is fundamental
in the proofs of the theorems; given that whenever the inverse matrix
exists, it is unique, it will not be necessary to explain the calculation
leading to it: it will be enough to check that the proposed matrix
(in the corresponding theorem) is the inverse matrix one was looking
for. The matrices appearing during the demostration process will not
be circulant; therefore, usual analytical inversion processes of circulant
matrix inversion will not be valid.

This paper is organized as follows. First, synchronized states will
be considered, this solution being quite straightforward, it will
indicate how to face up to the more complicated travelling waves in
the next section. Both results will be used to study the period-doubling
cascades of the patterns. The paper concludes with a section indicating
connections of this work with other researchs.

\section{Regular and chaotic synchronization}

In this section straightforward analytical results will be presented
for synchronization in CML, that is, for all the oscilators having
the same value at anytime. This is a striking behaviour, in particular
when chaotic syncronization is produced, where chaotic systems are
very sensitive to perturbations and it is supposed that any slight
modification generated by the coupling of the oscillators of CML would
destroy the synchronization. The mathematical approach to this problem
is far from being unique \cite{Anteneodo}.

~

Let \begin{equation}
X_{i}(n+1)=(1-\alpha)f(X_{i}(n))+\frac{\alpha}{m}\sum_{i=1}^{m}f(X_{i}(n))\:\;\quad i=1,\,\dots,\, m\label{eq:tres}\end{equation}
be the CML, with $m$ oscillators, being $\alpha$ the coupling parameter
and $f(x)$ a function depending on a parameter $r$, in function
of which the system $y_{n+1}=f(y_{n};r)$ shows fixed points for some
arbitrary period $p$.

\subsection{Fixed points of the system. Stationary synchronized state}

It is straightforward to get the fixed points of the system. If the
function $f(x)$ has a fixed point in $x^{*}$ then $(x^{*},x^{*},\dots^{m)},x^{*})$
will be a fixed point of the system given by (\ref{eq:tres}), since
if\[
X_{i}(n)=x^{*}\quad i=1,\,\dots,\, m\]
it turns out that\[
f(X_{i}(n))=f(x^{*})=x^{*}=X_{i}(n)\quad i=1,\,\dots,\, m\]
and\[
\begin{array}{c}
X_{i}(n+1)=(1-\alpha)f(x^{*})+{\displaystyle \frac{\alpha}{m}\sum_{i=1}^{m}}f(x^{*})=f(x^{*})=x^{*}\\
i=1,...,m\end{array}\]
as it was wanted to prove.

It is then deduced that:\begin{equation}
X(n)=(x^{*},x^{*},...^{m)},x^{*})\label{eq:dos}\end{equation}
is a stationary sychronized state of the system. 

\begin{figure}
\begin{centering}
\includegraphics[angle=-90,width=0.8\textwidth]{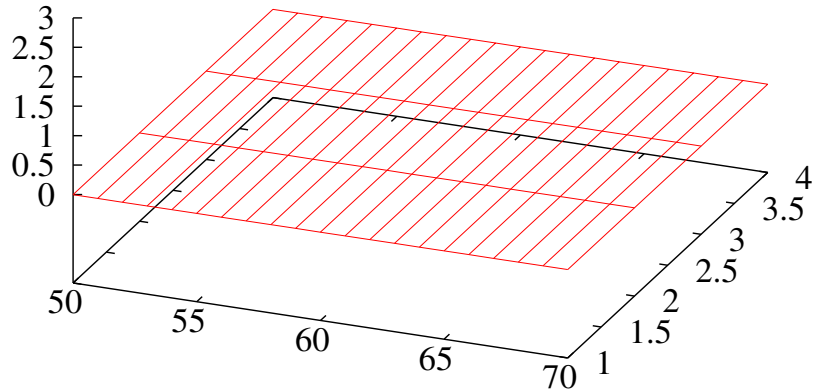}
\par\end{centering}

\caption{\label{fig:SSS1}Stationary synchronized state. CML with $f(x;r)=rx(1-x)$,
$r=1.0$, $\alpha=0.1$ and $\varepsilon=0.1$}

\end{figure}

\begin{figure}
\begin{centering}
\includegraphics[angle=-90,width=0.8\textwidth]{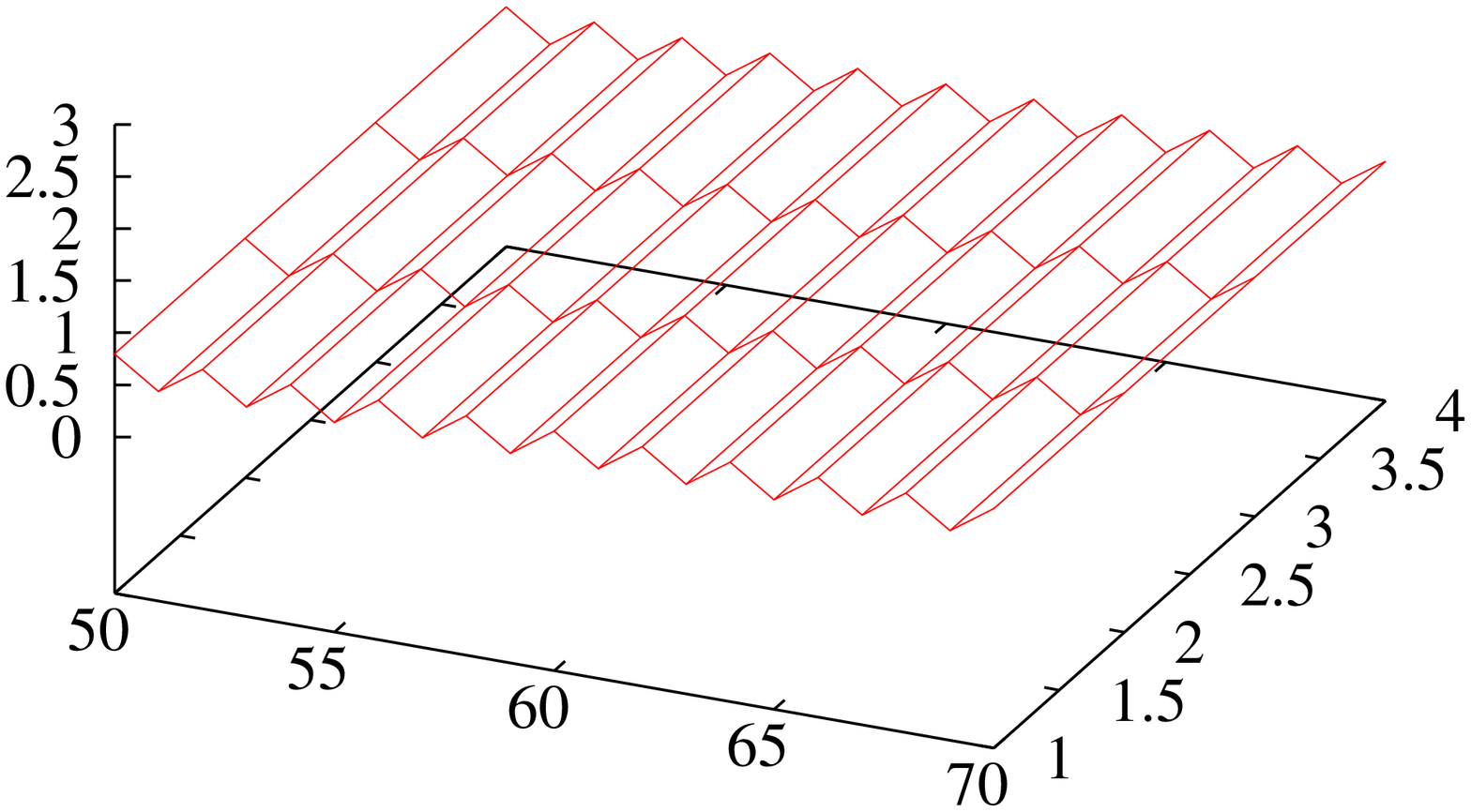}
\par\end{centering}

\caption{\label{fig:SSS2}Period-$2$ synchronized state. CML with $f(x;r)=rx(1-x)$,
$r=3.2$, $\alpha=0.1$ and $\varepsilon=0.1$}

\end{figure}

\begin{figure}
\begin{centering}
\includegraphics[angle=-90,width=0.8\textwidth]{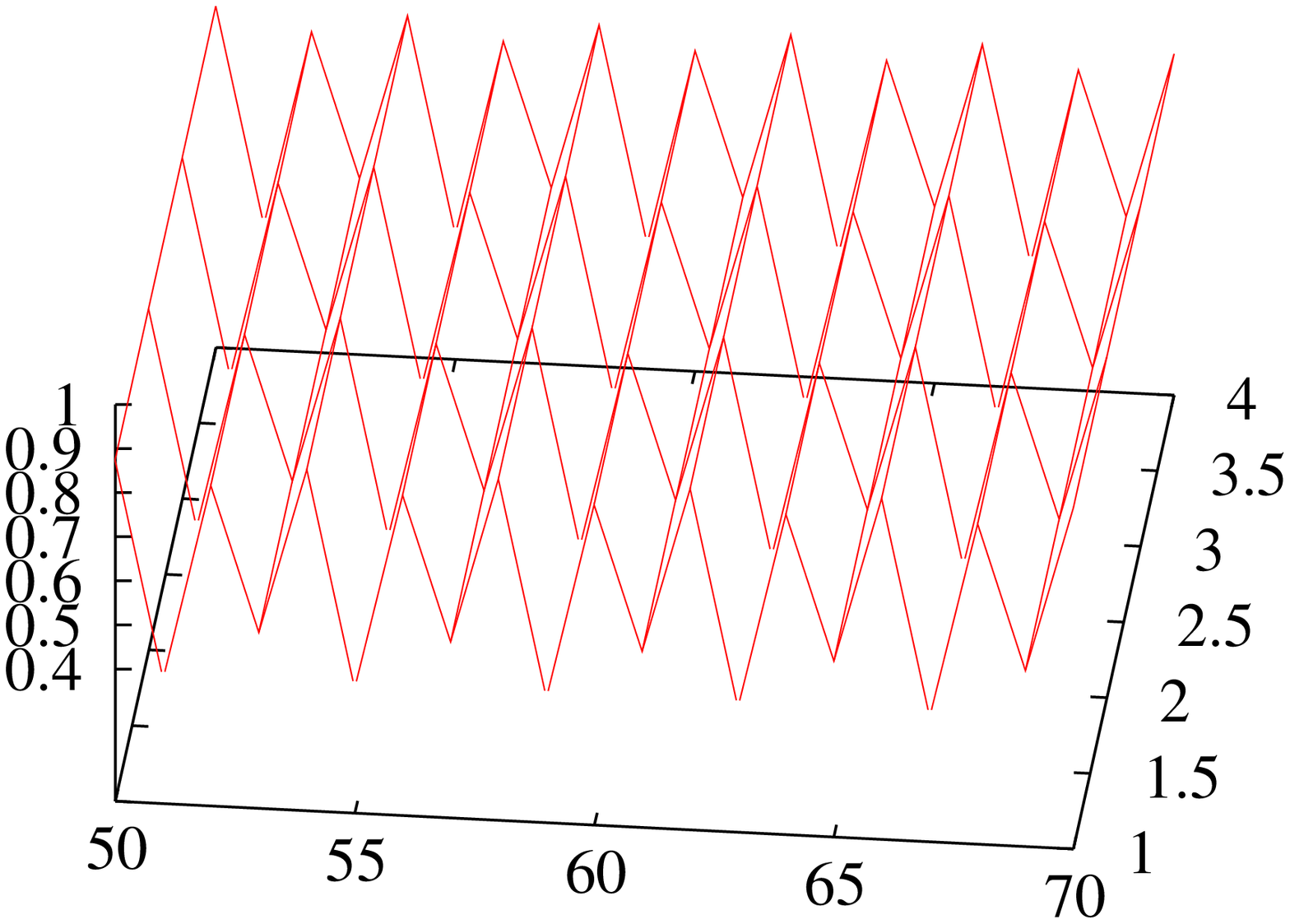}
\par\end{centering}

\caption{\label{fig:SSS3}Period-$4$ synchronized state. CML with $f(x;r)=rx(1-x)$,
$r=3.4985$, $\alpha=0.1$ and $\varepsilon=0.1$}

\end{figure}

\begin{figure}
\begin{centering}
\includegraphics[angle=-90,width=0.8\textwidth]{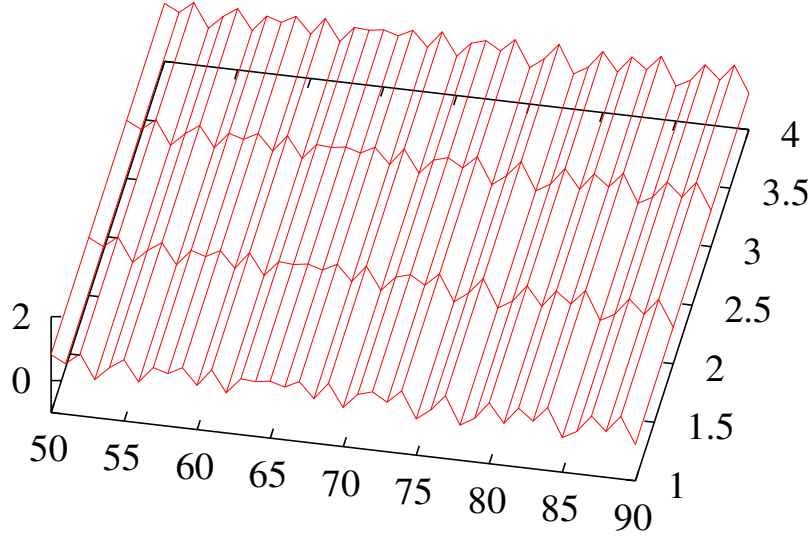}
\par\end{centering}

\caption{\label{fig:SSS4}Chaotic synchronized state. CML with $f(x;r)=rx(1-x)$,
$r=3.9$, $\alpha=0.1$ and $\varepsilon=0.1$}

\end{figure}

It is observed that if chosen $x^{*}\;\mbox{and}\: r$ for $f(x^{*};r)$
to determine a periodical or chaotic evolution of $x^{*}$, then the
result would be that CML would have correspondingly periodical or
chaotic synchronization, being this a proof of the existence of synchronized
states both periodical and chaotic. In contrast from the CML with
nearest neighbour coupling, there is not an upper limit in the number
of oscillators {}``$m$'' such that stable synchronous chaotic state
exists \cite{Bohr1993}. See figures \ref{fig:SSS1}-\ref{fig:SSS4}.

~

Let us now study the linear stability of fixed points, where the eigenvalues
of jacobian matrix will be calculated. 

The jacobian matrix is given by:

\[
\left(\frac{\partial X_{i}(n+1)}{\partial X_{j}(n)}\right)_{X^{*}}=\left(\begin{array}{cccc}
(1-\frac{m-1}{m}\alpha) & \frac{\alpha}{m} & \cdots & \frac{\alpha}{m}\\
\frac{\alpha}{m} & (1-\frac{m-1}{m}\alpha) & \cdots & \frac{\alpha}{m}\\
\vdots & \vdots & \ddots & \vdots\\
\frac{\alpha}{m} & \frac{\alpha}{m} & \cdots & (1-\frac{m-1}{m}\alpha)\end{array}\right)f^{\prime}(x^{*})\]
where $X^{*}=\left(x^{*},x^{*},...,x^{*}\right)$ and its eigenvalues
are

\[
\left\{ \begin{array}{cc}
\lambda=f^{\prime}(x^{*}) & \mbox{single}\\
\lambda=(1-\alpha)f^{\prime}(x^{*}) & \mbox{ multiplicity }(m-1)\end{array}\right.\]
therefore, the fixed point given by (\ref{eq:dos}), or what would
be the same, the stationary schronized state, would be stable whenever
$x^{*}$ is a stable fixed point of $f(x)$.

\subsection{Periodical synchronized states}

The existence of periodical syncronized states is reflected in theorem
1, shown below. The way to proceed with this proof is similar to the
one used to obtain the stationary synchronized state.

\begin{description}
\item [{Theorem~1.}] Let $\left\{ x_{1}^{*},x_{2}^{*},...,x_{p}^{*}\right\} $
be a $p$-periodic orbit of $C^{1}$ function $f$. Then the CML given
by \[
X_{i}(n+1)=(1-\alpha)f(X_{i}(n))+\frac{\alpha}{m}\sum_{i=1}^{m}f(X_{i}(n))\qquad i=1,\dots,\, m\]

\begin{enumerate}
\item [i)] shows a synchronized state of the same period as the function
$f$. The sychronized states are as follows \[
(x_{j}^{*},x_{j}^{*},\ldots^{m)},x_{j}^{*})_{j=1,\,\dots,\, p}\]

\item [ii)] the sychronized states $(x_{j}^{*},x_{j}^{*},\ldots^{m)},x_{j}^{*})_{j=1,\,\dots,\, p}$
have the same stability as the fixed point $x_{j}^{*}$ of $f^{p}$,
$j=1,\,\dots,\, p$.
\end{enumerate}
\item [{Proof}]~

\begin{enumerate}
\item [i)] Taking, for a given time $n$,\[
X_{i}(n)=x_{1}^{*}\quad i=1,\,\dots,\, m\]
{\large{} }results in\[
f^{p}(X_{i}(n))=X_{i}(n)\quad i=1,\,\dots,\, m\]
with the first iteration of CML being: \[
X_{i}(n+1)=(1-\alpha)f(x_{1}^{*})+\frac{\alpha}{m}\sum_{j=1}^{m}f(x_{1}^{*})=f(x_{1}^{*})\qquad i=1,\,\dots,\, m\]
and the $p$-th iteration being:\[
\begin{array}{rl}
X_{i}(n+p)= & (1-\alpha)f(X_{i}(n+p-1))+{\displaystyle \frac{\alpha}{m}}{\displaystyle \sum_{j=1}^{m}}f(X_{j}(n+p-1))\\
= & f^{p}(x_{1}^{*})=x_{1}^{*}\qquad i=1,\,...,\, m\end{array}\]
As a result, the CML shows $p$ fixed points $(x_{j}^{*},x_{j}^{*},\cdots x_{j}^{*})_{j=1,\cdots,p}$
of period $p$, which constitute one sychronized state of the same
period. These orbits of period $p$ constitute patterns of the CML.
\item [ii)] To study the stability of the fixed points $(x_{j}^{*},x_{j}^{*},\cdots x_{j}^{*})_{j=1,\cdots,p}$
it is enough to perform it in $X_{1}^{*}=(x_{1}^{*},x_{1}^{*},\cdots x_{1}^{*})$,
because $f^{p^{\prime}}(x_{j}^{*})$ has the same value for every
fixed point $x_{j}^{*}$ of the $p$-periodic orbit.

Let us calculate the eigenvalues of the jacobian matrix of the $p$-th
iterate in that point. 

To calculate this Jacobian matrix, one must observe the following,
applying the chain rule: \[
\begin{array}{l}
\left(\frac{\partial X_{i}(n+p)}{\partial X_{j}(n)}\right)_{X_{1}^{*}}=\\
\\\qquad\left(\begin{array}{cccc}
(1-\frac{m-1}{m}\alpha) & \frac{\alpha}{m} & \cdots & \frac{\alpha}{m}\\
\frac{\alpha}{m} & (1-\frac{m-1}{m}\alpha) & \cdots & \frac{\alpha}{m}\\
\vdots & \vdots & \ddots & \vdots\\
\frac{\alpha}{m} & \frac{\alpha}{m} & \cdots & (1-\frac{m-1}{m}\alpha)\end{array}\right)f^{\prime}(x_{p}^{*})\left(\frac{\partial X_{i}(n+p-1)}{\partial X_{j}(n)}\right)_{X_{1}^{*}}\end{array}\]

finally as a result:\[
\begin{array}{l}
\left(\frac{\partial X_{i}(n+p)}{\partial X_{j}(n)}\right)_{X_{1}^{*}}=\\
\\\:\left(\begin{array}{cccc}
\frac{1}{m}+\frac{m-1}{m}(1-\alpha)^{p} & \frac{1}{m}-\frac{1}{m}(1-\alpha)^{p} & \cdots & \frac{1}{m}-\frac{1}{m}(1-\alpha)^{p}\\
\frac{1}{m}-\frac{1}{m}(1-\alpha)^{p} & \frac{1}{m}+\frac{m-1}{m}(1-\alpha)^{p} & \cdots & \frac{1}{m}-\frac{1}{m}(1-\alpha)^{p}\\
\vdots & \vdots & \ddots & \vdots\\
\frac{1}{m}-\frac{1}{m}(1-\alpha)^{p} & \frac{1}{m}-\frac{1}{m}(1-\alpha)^{p} & \cdots & \frac{1}{m}+\frac{m-1}{m}(1-\alpha)^{p}\end{array}\right)\prod_{i=1}^{p}f^{\prime}(x_{i}^{*})\end{array}\]
 \[
\]
with eigenvalues \begin{equation}
\left\{ \begin{array}{cc}
\lambda={\displaystyle \prod_{i=1}^{p}}f^{\prime}(x_{i}^{*}))=f^{p\prime}(x_{1}^{*}) & \mbox{single}\\
\lambda=(1-\alpha)^{p}{\displaystyle \prod_{i=1}^{p}}f^{\prime}(x_{i}^{*})=(1-\alpha)^{p}f^{p\prime}(x_{1}^{*}) & \mbox{ multiplicity }(m-1)\end{array}\right.\label{eq:cuatro}\end{equation}

So, $(x_{1}^{*},\cdots,x_{1}^{*})$ is a stable point of the CML,
and therefore, it is in a stable sychronized state of period $p$,
whenever $x_{1}^{*}$ is the stable fixed point $f^{p}$. Furthermore,
as $f^{p\prime}(x_{1}^{*})=f^{p\prime}(x_{j}^{*})\quad j=1,\dots,\, m$
all points have the same stability.

\end{enumerate}
\end{description}
Keep in mind that if in \textbf{Theorem 1} $p=1$, then the stationary
sychronized state previously studied is recovered; and because of
this, it will undergo the period-doubling process that will be described
in what follows.

\subsection{Period doubling cascade of periodic synchronized states}

One would expect that if the function $f^{p}$, from \textbf{Theorem
1}, undergoes a period doubling cascade, then the CML given by (\ref{eq:tres})
shows a duplication cascade in the sychronized states of period $p$
derived in \textbf{Theorem 1}. 

\begin{figure}
\begin{centering}
\includegraphics[angle=-90,width=0.8\textwidth]{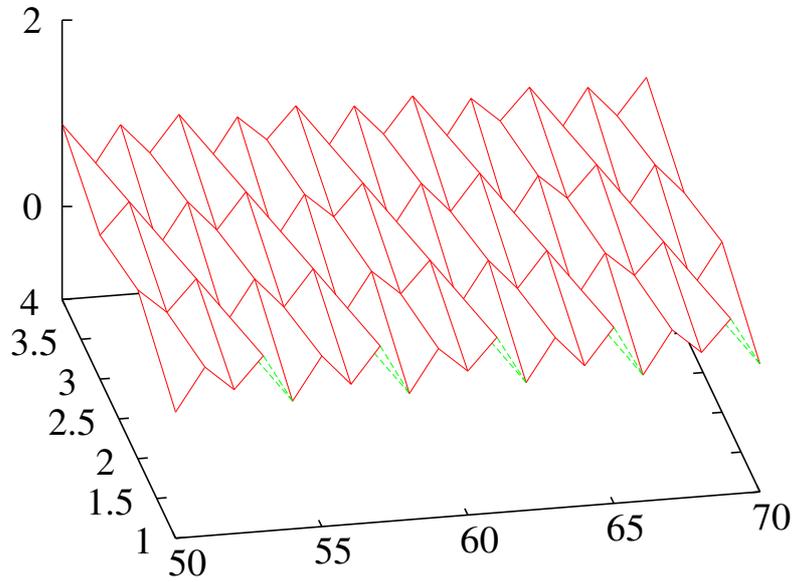}
\par\end{centering}

\caption{\label{fig:p4}Period-$4$ travelling wave. CML with $f(x;r)=rx(1-x)$,
$r=3.55464$, $\alpha=0.1$ and $\varepsilon=0.001$}

\end{figure}

\begin{figure}
\begin{centering}
\includegraphics[angle=-90,width=0.8\textwidth]{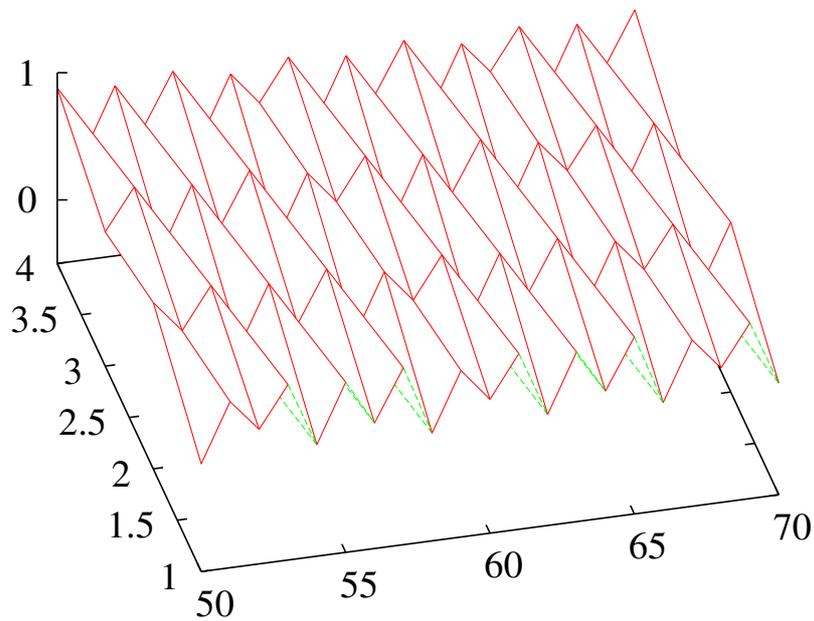}
\par\end{centering}

\caption{\label{fig:p8}Period-$8$ travelling wave. CML with $f(x;r)=rx(1-x)$,
$r=3.566667$, $\alpha=0.1$ and $\varepsilon=0.001$}

\end{figure}

\begin{description}
\item [{Theorem~2.}] The synchronized states of period $p$ given by Theorem
1 undergo a period-doubling cascade as does $f^{p}$.
\item [{Proof}]~

The proof is straightforward using Theorem 1, simply using successive
substitution of $p$ by $p\cdot2,\, p\cdot2^{2},\,\dots,\, p\cdot2^{n},\,\dots$
every time that the $p$-periodic orbit of $f$ undergoes a period
doubling bifurcation according to that theorem.

\item [{Note:}] See figures \ref{fig:p4} and \ref{fig:p8}.
\end{description}

\subsection{A nonexistence theorem}

It has been proven, in Theorem 1, the existences of $p$-period synchronized
states in the CML, formed by the points of $p$-periodic orbit of
$f$.

One may ask whether the $n$-tuple of the form

\[
\begin{array}{rl}
(X_{1}(n),X_{2}(n),\cdots,X_{p}(n))= & (x_{j}^{*},f(x_{j}^{*}),...,f^{p-1}(x_{j}^{*}))\\
f^{p}(x_{j}^{*})= & x_{j}^{*}\quad j=1,\cdots p\end{array}\]
that indicates that every oscillator is positioned in the successive
points of the $p$-periodic orbit, generates a pattern of period $p$
in the CML; that is to say, a travelling wave. Nevertheless, this
presumption is false, as shown below. 

\begin{description}
\item [{Theorem~3.}] Let $\left\{ x_{1}^{*},x_{2}^{*},...,x_{p}^{*}\right\} $
be a $p$-periodic orbit of the function $f$, then the CML given
by \[
X_{i}(n+1)=(1-\alpha)f(X_{i}(n-1))+\frac{\alpha}{p}\sum_{i=1}^{p}f(X_{i}(n-1))\qquad i=1,\,\dots,\, p\]
does not have a $p$-periodic orbit of the form \[
(X_{1}(n),X_{2}(n),\cdots,X_{p}(n))=(x_{j}^{*},f(x_{j}^{*}),...,f^{p-1}(x_{j}^{*}))\]
$x_{j}^{*}$ being any of the $p$ points of the $p$-periodic orbit.
\item [{Proof}]~

The following initial conditions are taken\[
(X_{1}(n),X_{2}(n),\cdots,X_{p}(n))=(x_{1}^{*},x_{2}^{*},...,x_{p}^{*})\]
After the first iteration, it will become\[
\left\{ \begin{array}{c}
X_{1}(n+1)=x_{2}^{*}=(1-\alpha)f(x_{1}^{*})+\frac{\alpha}{p}(f(x_{1}^{*})+...+f(x_{p}^{*}))\\
X_{2}(n+1)=x_{3}^{*}=(1-\alpha)f(x_{2}^{*})+\frac{\alpha}{p}(f(x_{1}^{*})+...+f(x_{p}^{*}))\\
\cdots\\
X_{p}(n+1)=x_{1}^{*}=(1-\alpha)f(x_{p}^{*})+\frac{\alpha}{p}(f(x_{1}^{*})+...+f(x_{p}^{*}))\end{array}\right.\]
therefore:\[
\left\{ \begin{array}{c}
x_{2}^{*}=(1-\alpha)x_{2}^{*}+\frac{\alpha}{p}(x_{1}^{*}+...+x_{p}^{*})\\
x_{3}^{*}=(1-\alpha)x_{3}^{*}+\frac{\alpha}{p}(x_{1}^{*}+...+x_{p}^{*})\\
\cdots\\
x_{1}^{*}=(1-\alpha)x_{1}^{*}+\frac{\alpha}{p}(x_{1}^{*}+...+x_{p}^{*})\end{array}\right.\]
operating it results in:\[
\left\{ \begin{array}{c}
\alpha x_{2}^{*}=\frac{\alpha}{p}(x_{1}^{*}+...+x_{p}^{*})\\
\alpha x_{3}^{*}=\frac{\alpha}{p}(x_{1}^{*}+...+x_{p}^{*})\\
\cdots\\
\alpha x_{1}^{*}=\frac{\alpha}{p}(x_{1}^{*}+...+x_{p}^{*})\end{array}\right.\]
from which it is deduced that\[
x_{1}^{*}=x_{2}^{*}=...=x_{p}^{*}\]
in contradiction with $x_{1}^{*}\neq x_{2}^{*}\neq...\neq x_{p}^{*}$.

\end{description}
This negative result, about $p$-periodic waves, brings us to question
the conditions under which they are produced. This study is conducted
in the following section.

~

\section{Analytical study of patterns in weakly coupled CML}

\subsection{Travelling waves}

It is has been proven in Theorem 3 that the $p$-periodic orbit of
the function $f(x;r)$ is not inherited by the system, but it is easily
observed that if $\alpha=0$ then a wave of this period exists in
the CML. Given that for $\alpha=0$ the wave exists, one would wonder
if for a small coupling $\alpha\ll1$, the CML admits a pertubative
solution. For this study, we will substitute $\alpha$ with $\varepsilon\alpha$,
having $\varepsilon\ll1$ and assuming the new $\alpha$ is $O(1)$,
in the CML given by (\ref{eq:uno}).

\begin{description}
\item [{Theorem~4.}] Let $\left\{ x_{1}^{*},x_{2}^{*},...,x_{p}^{*}\right\} $
be a $p$-periodic orbit of a $C^{2}$ function $f$, such that $f^{p\prime}(x_{i}^{*})\neq1$,
$i=1,\,\dots,\, p$, then the CML given by \begin{equation}
\begin{array}{c}
X_{i}(n+1)=(1-\varepsilon\alpha)f(X_{i}(n))+{\displaystyle \frac{\alpha\varepsilon}{p}\sum_{j=1}^{p}}f(X_{j}(n))\\
i=1,\,\dots,\, p\quad\varepsilon\ll1\end{array}\label{eq:cinco}\end{equation}
shows a $p$-periodic solution given by\[
\begin{array}{c}
X_{i}(n+j)=x_{i+j}^{*}+\varepsilon A_{i+j}\\
\begin{array}{c}
i=1,\,\dots,\, p\\
j=0,\,\dots,\, p-1\end{array}\end{array}\]
where\[
A_{k}=\frac{\alpha}{(1-(f^{p}(x_{1}))^{\prime})}\sum_{j=1}^{p}\left[f^{p-j+k-1}(x_{j+1}^{*})\right]^{\prime}\left(\left(x_{j+1}^{*}\right)+\frac{1}{p}\sum_{l=1}^{p}x_{l}^{*}\right)\qquad k=1,\dots,p\]
with periodic conditions\[
\begin{array}{c}
A_{i+p}=A_{i}\\
x_{i+p}^{*}=x_{i}^{*}\end{array}\quad i=1,\,\dots,\, p\]

\item [{Proof}]~

The periodic orbit given by \[
\begin{array}{c}
X_{i}(n+j)=x_{i+j}^{*}+\varepsilon A_{i+j}\\
\begin{array}{c}
i=1,\dots,p\\
j=0,\dots,p-1\end{array}\end{array}\]
 will exist when the following system\begin{equation}
\left\{ \begin{array}{c}
X_{i}(n)=x_{i}^{*}+\epsilon A_{i}\\
X_{i}(n+1)=x_{i+1}^{*}+\epsilon A_{i+1}\\
\vdots\\
X_{i}(n+p-1)=x_{i+p-1}^{*}+\epsilon A_{i+p-1}=x_{i-1}^{*}+\epsilon A_{i-1}\\
X_{i}(n+p)=x_{i}^{*}+\epsilon A_{i}\end{array}\right.\, i=1,...,p\label{eq:seis}\end{equation}
is compatible and determined.

As\[
X_{i}(n+1)=(1-\varepsilon\alpha)f(X_{i}(n))+\frac{\epsilon\alpha}{p}\sum_{j=1}^{p}f(X_{j}(n))\]
from (\ref{eq:seis}), it reults in\begin{equation}
x_{i+1}^{*}+\varepsilon A_{i+1}=(1-\varepsilon\alpha)f(x_{i}^{*}+\varepsilon A_{i})+\frac{\varepsilon\alpha}{p}\sum_{j=1}^{p}f(x_{j}^{*}+\varepsilon A_{j})\label{siete}\end{equation}
Performing the expansion\[
f(x_{i}^{*}+\varepsilon A_{i})=f(x_{i}^{*})+\varepsilon A_{i}f^{\prime}(x_{i}^{*})+O(\varepsilon^{2})\]
and replacing in (\ref{siete}), the system results in\[
\begin{array}{c}
x_{i+1}^{*}+\varepsilon A_{i+1}=x_{i+1}^{*}+\varepsilon A_{i}f^{\prime}(x_{i}^{*})-\varepsilon\alpha x_{i+1}^{*}+{\displaystyle \frac{\varepsilon\alpha}{p}\sum_{j=1}^{p}}(x_{j+1}^{*})+O(\varepsilon^{2})\\
i=1,...,p\end{array}\]
Solving the system to order $\varepsilon$ it is obtained:\[
-A_{i}f^{\prime}(x_{i}^{*})+A_{i+1}=\alpha x_{i+1}^{*}+\frac{\alpha}{p}\sum_{j=1}^{p}x_{j+1}^{*}\qquad i=1,\dots,p\]
 whose matricial expression is:{\scriptsize \begin{equation}
\left(\begin{array}{cccccc}
-f^{\prime}(x_{1}^{*}) & 1 & 0 & 0 & \cdots & 0\\
0 & -f^{\prime}(x_{2}^{*}) & 1 & 0 & \cdots & 0\\
0 & 0 & -f^{\prime}(x_{3}^{*}) & 1 & \cdots & 0\\
\vdots & \vdots & \vdots & \vdots & \ddots & \vdots\\
1 & 0 & 0 & 0 & \cdots & -f^{\prime}(x_{p}^{*})\end{array}\right)\left(\begin{array}{c}
A_{1}\\
A_{2}\\
A_{3}\\
\vdots\\
A_{p}\end{array}\right)=\alpha\left(\begin{array}{c}
-x_{2}^{*}+\frac{1}{p}\Sigma_{j=1}^{p}x_{j}^{*}\\
-x_{3}^{*}+\frac{1}{p}\Sigma_{j=1}^{p}x_{j}^{*}\\
-x_{4}^{*}+\frac{1}{p}\Sigma_{j=1}^{p}x_{j}^{*}\\
\vdots\\
-x_{1}^{*}+\frac{1}{p}\Sigma_{j=1}^{p}x_{j}^{*}\end{array}\right)\label{nueve}\end{equation}
}{\scriptsize \par}

This is a system of $p$ equations and $p$ unknowns whose coefficient
matrix has determinant

\[
(-1)^{p}\prod_{i=1}^{p}f^{\prime}(x_{i}^{*})+(-1)^{p+1}\]

Given that $\prod_{i=1}^{p}f^{\prime}(x_{i}^{*})=f^{p\prime}(x_{i}^{*})\neq1$
(by hypothesis) the system is compatible and determined for every
$\alpha\neq0$. Moreover, the solution of the system is different
from the trivial one, since the independent term column is not null
$x_{1}^{*}\neq x_{2}^{*}\neq\cdots\neq x_{p}^{*}$, that is\[
\left(\begin{array}{c}
-x_{2}^{*}+{\displaystyle \frac{1}{p}\sum_{j=1}^{p}}x_{j}^{*}\\
-x_{3}^{*}+{\displaystyle \frac{1}{p}\sum_{j=1}^{p}}x_{j}^{*}\\
-x_{4}^{*}+{\displaystyle \frac{1}{p}\sum_{j=1}^{p}}x_{j}^{*}\\
\vdots\\
-x_{1}^{*}+{\displaystyle \frac{1}{p}\sum_{j=1}^{p}}x_{j}^{*}\end{array}\right)\neq\left(\begin{array}{c}
0\\
0\\
0\\
\vdots\\
0\end{array}\right)\]

\end{description}
~

It does not matter which oscillator is considered for study of the
evolution of the system, as the algebraic system obtained is always
the same.

The solution of the system in (\ref{nueve}) can be obtained directly
by inversion and results in:

\[
\left(\begin{array}{c}
A_{1}\\
A_{2}\\
\vdots\\
A_{p}\end{array}\right)=\alpha\left(\begin{array}{ccccc}
-f^{\prime}(x_{1}^{*}) & 1 & 0 & \cdots & 0\\
0 & -f^{\prime}(x_{2}^{*}) & 1 & \cdots & 0\\
\vdots & \vdots & \vdots & \ddots & \vdots\\
1 & 0 & 0 & \cdots & -f^{\prime}(x_{p}^{*})\end{array}\right)^{-1}\left(\begin{array}{c}
-x_{2}^{*}+{\displaystyle \frac{1}{p}\sum_{j=1}^{p}}x_{j}^{*}\\
-x_{3}^{*}+{\displaystyle \frac{1}{p}\sum_{j=1}^{p}}x_{j}^{*}\\
\vdots\\
-x_{1}^{*}+{\displaystyle \frac{1}{p}\sum_{j=1}^{p}}x_{j}^{*}\end{array}\right)\]

The inversion of the matrix (which is not a circulant one) results
in the following\begin{equation}
\left(\begin{array}{c}
A_{1}\\
A_{2}\\
A_{3}\\
\vdots\\
A_{p}\end{array}\right)=\alpha\frac{1}{(-1)^{p+1}(1-(f^{p}(x_{1}^{*}))^{\prime})}MN\label{eq:diez}\end{equation}
where the matrix $M$ is given by

{\footnotesize ~}{\footnotesize \par}

{\scriptsize \begin{equation}
M=\left(\begin{array}{ccccc}
f^{\prime}(x_{2}^{*})\cdots f^{\prime}(x_{p}^{*}) & f^{\prime}(x_{3}^{*})\cdots f^{\prime}(x_{p}^{*}) & f^{\prime}(x_{4}^{*})\cdots f^{\prime}(x_{p}^{*}) & \cdots & 1\\
1 & f^{\prime}(x_{3}^{*})\cdots f^{\prime}(x_{p}^{*})f^{\prime}(x_{1}^{*}) & f^{\prime}(x_{4}^{*})\cdots f^{\prime}(x_{p}^{*})f^{\prime}(x_{1}^{*}) & \cdots & f^{\prime}(x_{1}^{*})\\
f^{\prime}(x_{2}) & 1 & f^{\prime}(x_{4}^{*})\cdots f^{\prime}(x_{p}^{*})f^{\prime}(x_{1}^{*})f^{\prime}(x_{2}^{*}) & \cdots & f^{\prime}(x_{1}^{*})f^{\prime}(x_{2}^{*})\\
\vdots & \vdots & \vdots & \ddots & \vdots\\
f^{\prime}(x_{2}^{*})\cdots f^{\prime}(x_{p-1}^{*}) & f^{\prime}(x_{3}^{*})\cdots f^{\prime}(x_{p-1}^{*}) & f^{\prime}(x_{4}^{*})\cdots f^{\prime}(x_{p-1}^{*}) & \cdots & f^{\prime}(x_{1}^{*})f^{\prime}(x_{2}^{*})\cdots f^{\prime}(x_{p-1}^{*})\end{array}\right)\label{eq:matrizM}\end{equation}
}and $N$ by

\[
N=\left(\begin{array}{c}
-x_{2}^{*}+{\displaystyle \frac{1}{p}\sum_{j=1}^{p}}x_{j}^{*}\\
-x_{3}^{*}+{\displaystyle \frac{1}{p}\sum_{j=1}^{p}}x_{j}^{*}\\
-x_{4}^{*}+{\displaystyle \frac{1}{p}\sum_{j=1}^{p}}x_{j}^{*}\\
\vdots\\
-x_{1}^{*}+{\displaystyle \frac{1}{p}\sum_{j=1}^{p}}x_{j}^{*}\end{array}\right)\]
~

After operating in (\ref{eq:diez}) it results in:

\begin{equation}
\begin{array}{c}
A_{k}=\frac{\alpha}{(-1)^{p+1}(1-(f^{p}(x_{1}))^{\prime})}{\displaystyle \sum_{j=1}^{p}}\left[f^{p-j+k-1}(x_{j+1}^{*})\right]^{\prime}\left(\left(x_{1+j}^{*}\right)+{\displaystyle \frac{1}{p}\sum_{l=1}^{p}}x_{l}^{*}\right)\\
k=1,...,p\end{array}\label{soluc}\end{equation}

Every $A_{k}\neq0$ because the solution is known to be different
from the trivial one.

The solution obtained is valid at order $O(\varepsilon^{2})$ while
$\varepsilon\ll\frac{1}{1-f^{p\prime}(x_{1}^{*})}$.

\subsection{Period doubling cascade for travelling waves in a CML}

Period-doubling transitions to chaos have already been observed a
long time ago, in CML with nearest neighbour coupling, using the Mandelbrot
map \cite{Selection1992}. The existence of this phenomenon is not
relegated only to the quadratic functions, and its existence can be
proved for any function (as we will demonstrate) undergoing a period-doubling
cascade; therefore, this phenomenon must be very frequent.

\begin{description}
\item [{Theorem~5.}] Let $f:I\rightarrow I$ be a $C^{2}$ funtion depending
on some parameter, in function of which the $2^{p}$-periodic orbit
of the map $x_{n+1}=f(x_{n})$ undergoes a period-doubling cascade.
Let $\left\{ x_{i,2^{p+q}}^{*}\right\} _{i=1}^{2^{p+q}}$ be the $2^{p+q}$-period
orbit of the cascade, $q\in\mathbb{N}$, where it is noted that $f^{k}(x_{i,2^{p+q}}^{*})=x_{i+k,2^{p+q}}^{*}$.

The CML given by \begin{equation}
\begin{array}{rl}
X_{i}(n+1) & =(1-\varepsilon\alpha)f(X_{i}(n))+{\displaystyle \frac{\alpha\varepsilon}{2^{p}}\sum_{j=1}^{2^{p}}}f(X_{j}(n))\\
i & =1,\cdots,2^{p}\quad\varepsilon\ll1\end{array}\label{eq:ONCE}\end{equation}
has a $2^{p+q}$-periodic solution given by\[
\begin{array}{rl}
X_{i}(n+j)= & x_{2^{q}(i-1)+1+j,2^{p+q}}^{*}+\varepsilon A_{2^{q}(i-1)+1+j}\\
i= & 1,\dots,2^{p}\quad j=0,\dots,2^{p+q}\end{array}\]
where \[
\begin{array}{c}
A_{k}=\frac{\alpha}{(-1+(f^{2^{p+q}}(x_{1}^{*}))^{\prime})}{\displaystyle \sum_{j=1}^{2^{p+q}}}\left[f^{2^{p+q}-j+k-1}(x_{j+1,2^{p+q}}^{*})\right]^{\prime}\left(\left(x_{1+j,2^{p+q}}^{*}\right)+\frac{1}{p}S_{j}\right)\\
k=1,\dots,2^{p+q}\end{array}\]
with\[
S_{j}=\left\{ \begin{array}{lcl}
{\displaystyle \sum_{i=1}^{2^{p}}}x_{2^{q}(i-1)+2,2^{p+q}}^{*} & \quad & \mbox{if}\quad j=[2^{q}]\\
{\displaystyle \sum_{i=1}^{2^{p}}}x_{2^{q}(i-1)+3,2^{p+q}}^{*} & \quad & \mbox{if}\quad j=[2^{q}]+1\\
{\displaystyle \sum_{i=1}^{2^{p}}}x_{2^{q}(i-1)+4,2^{p+q}}^{*} & \quad & \mbox{if}\quad j=[2^{q}]+2\\
{\displaystyle \sum_{i=1}^{2^{p}}}x_{2^{q}(i-1)+5,2^{p+q}}^{*} & \quad & \mbox{if}\quad j=[2^{q}]+3\\
\qquad\qquad\vdots & \vdots & \qquad\quad\vdots\\
{\displaystyle \sum_{i=1}^{2^{p}}}x_{2^{q}(i-1)+2^{q}+1,2^{p+q}}^{*} & \quad & \mbox{if}\quad j=[2^{q}]+2^{q}-1\end{array}\right.\]
where $[2^{q}]$ represents a multiple of $2^{q}$. This $2^{p+q}$-periodic
solution fulfills the periodicity condition\[
\begin{array}{c}
A_{i+2^{p+q}}=A_{i}\\
x_{i+2^{p+q},2^{p+q}}^{*}=x_{i,2^{p+q}}^{*}\end{array}\]

\item [{Proof}]~

\textbf{Note: }Notice that although there are just $2^{p}$ oscillators
the periodic wave will have period $2^{p+q}$, therefore, the system
that is dealt with in the proof will have $2^{p+q}$ equations. 

Let $2^{p}$ oscillators be with the initial conditions given by \begin{equation}
X_{i}(n)=x_{2^{q}(i-1)+1,2^{p+q}}^{*}+\varepsilon A_{2^{q}(i-1)+1}\quad i=1,\cdots,2^{p}\label{eq:17}\end{equation}
Initial conditions are fixed points of $f^{2^{p+q}}$ (taken one every
$2^{q}$) plus a perturbation that must be calculated. A $2^{p+q}$-periodic
orbit will exist whenever the system{\footnotesize \begin{equation}
\left\{ \begin{array}{rl}
X_{i}(n)= & x_{2^{q}(i-1)+1,2^{p+q}}^{*}+\varepsilon A_{2^{q}(i-1)+1}\\
X_{i}(n+1)= & x_{2^{q}(i-1)+2,2^{p+q}}^{*}+\varepsilon A_{2^{q}(i-1)+2}\\
 & \vdots\\
X_{i}(n+2^{p+q})= & x_{2^{q}(i-1)+1+2^{p+q},2^{p+q}}^{*}+\varepsilon A_{2^{q}(i-1)+2^{p+q}}\\
= & x_{2^{q}(i-1)+1,2^{p+q}}^{*}+\varepsilon A_{2^{q}(i-1)+1}=X_{i}(n)\end{array}\right.\, i=1,...,2^{p}\label{eq:dieciocho}\end{equation}
}is compatible and determined.

From equation (\ref{eq:ONCE}) and subtituting the second equality
in (\ref{eq:dieciocho}), we have \begin{equation}
\begin{array}{c}
x_{2^{q}(i-1)+2,2^{p+q}}^{*}+\varepsilon A_{2^{q}(i-1)+2}=(1-\varepsilon\alpha)f(x_{2^{q}(i-1)+1,2^{p+q}}^{*}+\varepsilon A_{2^{q}(i-1)+1})+\\
+{\displaystyle \frac{\alpha\varepsilon}{2^{p}}\sum_{j=1}^{2^{p}}}f(x_{2^{q}(j-1)+1,2^{p+q}}^{*}+\varepsilon A_{2^{q}(j-1)+1})\end{array}\label{eq:diecinueve}\end{equation}

Performing a Taylor expansion of $f$ to order $O(\varepsilon^{2})$
and substituting in (\ref{eq:diecinueve}) the following is obtained\[
\begin{array}{c}
x_{2^{q}(i-1)+2,2^{p+q}}^{*}+\varepsilon A_{2^{q}(i-1)+2}=x_{2^{q}(i-1)+2,2^{p+q}}^{*}+\\
+\varepsilon A_{2^{q}(i-1)+1}f^{\prime}(x_{2^{q}(i-1)+1,2^{p+q}}^{*})-\varepsilon\alpha x_{2^{q}(i-1)+2,2^{p+q}}^{*}+\frac{\varepsilon\alpha}{2^{p}}\sum_{i=1}^{2^{p}}x_{2^{q}(i-1)+2,2^{p+q}}^{*}+O(\varepsilon^{2})\end{array}\]

Doing exactly the same with the next equality in (\ref{eq:dieciocho})
the following is obtained:\[
\begin{array}{c}
x_{2^{q}(i-1)+3,2^{p+q}}^{*}+\varepsilon A_{2^{q}(i-1)+3}=x_{2^{q}(i-1)+3,2^{p+q}}^{*}+\\
+\varepsilon A_{2^{q}(i-1)+2}f^{\prime}(x_{2^{q}(i-1)+2,2^{p+q}}^{*})-\varepsilon\alpha x_{2^{q}(i-1)+3,2^{p+q}}^{*}+{\displaystyle \frac{\varepsilon\alpha}{2^{p}}\sum_{i=1}^{2^{p}}}x_{2^{q}(i-1)+3,2^{p+q}}^{*}+O(\varepsilon^{2})\end{array}\]
and with the last equality in (\ref{eq:dieciocho}), we get the following
equation:\[
\begin{array}{c}
x_{2^{q}(i-1)+1,2^{p+q}}^{*}+\varepsilon A_{2^{q}(i-1)+1}=x_{2^{q}(i-1)+1,2^{p+q}}^{*}+\\
+\varepsilon A_{2^{q}(i-1)}f^{\prime}(x_{2^{q}(i-1),2^{p+q}}^{*})-\varepsilon\alpha x_{2^{q}(i-1)+1,2^{p+q}}^{*}+{\displaystyle \frac{\varepsilon\alpha}{2^{p}}\sum_{i=1}^{2^{p}}}x_{2^{q}(i-1)+1,2^{p+q}}^{*}+O(\varepsilon^{2})\end{array}\]
The former $2^{p+q}$-equations, for the oscillator i, represent a
linear system, whose matricial expresion is:{\scriptsize \begin{equation}
\begin{array}{c}
\left(\begin{array}{ccccc}
-f^{\prime}(x_{2^{q}(i-1)+1,2^{p+q}}^{*}) & 1 & 0 & \cdots & 0\\
0 & -f^{\prime}(x_{2^{q}(i-1)+2,2^{p+q}}^{*}) & 1 & \cdots & 0\\
\vdots & \vdots & \vdots & \ddots & \vdots\\
1 & 0 & 0 & \cdots & -f^{\prime}(x_{2^{q}(i-1)+2^{p+1},2^{p+q}}^{*})\end{array}\right).\left(\begin{array}{c}
A_{2^{q}(i-1)+1}\\
A_{2^{q}(i-1)+2}\\
\vdots\\
A_{2^{q}(i-1)+2^{p+q}}\end{array}\right)=\\
\\=\alpha\left(\begin{array}{c}
-x_{2^{q}(i-1)+2,2^{p+q}}^{*}+{\displaystyle \frac{1}{2^{p}}\sum_{i=1}^{2^{p+1}}}x_{2^{q}(i-1)+2,2^{p+q}}^{*}\\
-x_{2^{q}(i-1)+3,2^{p+q}}^{*}+{\displaystyle \frac{1}{2^{p}}\sum_{i=1}^{2^{p+1}}}x_{2^{q}(i-1)+3,2^{p+q}}^{*}\\
\vdots\\
-x_{2^{q}(i-1)+1,2^{p+q}}^{*}+{\displaystyle \frac{1}{2^{p}}\sum_{i=1}^{2^{p+1}}}x_{2^{q}(i-1)+1,2^{p+q}}^{*}\end{array}\right)\end{array}\label{quince}\end{equation}
}{\scriptsize \par}

The previous matricial expresion represents a system of $2^{p+q}$
equations with $2^{p+q}$ unknowns, and being the determinant of coefficient
matrix\[
(-1)^{2^{p+q}}\prod_{i=1}^{2^{p+q}}f^{\prime}(x_{i}^{*})-(-1)^{2^{p+q}}=\left[f^{2^{p+q}}(x_{i}^{*})\right]^{\prime}-1\neq0\]
(period doubling bifurcations take place when $\left[f^{2^{p+q}}(x_{i}^{*})\right]^{\prime}=-1$).

Thus the system is compatible and determined for every $\alpha$ and,
as in the previous cases, its solution is different from the trivial
one for $\alpha\neq0$.

The solution is obtained directly from (\ref{quince}) by inversion:\begin{equation}
\begin{array}{c}
\left(\begin{array}{c}
A_{2^{q}(i-1)+1}\\
A_{2^{q}(i-1)+2}\\
\vdots\\
A_{2^{q}(i-1)+2^{p+q}}\end{array}\right)=\frac{\alpha}{(-1+[f^{2^{p+q}}(x_{1}^{*})]^{\prime})}MN\end{array}\label{eq:veinte}\end{equation}
where\[
M=\left(\begin{array}{ccccc}
-f^{\prime}(x_{2^{q}(i-1)+1,2^{p+q}}^{*}) & 1 & 0 & \cdots & 0\\
0 & -f^{\prime}(x_{2^{q}(i-1)+2,2^{p+q}}^{*}) & 1 & \cdots & 0\\
\vdots & \vdots & \vdots & \ddots & \vdots\\
1 & 0 & 0 & \cdots & -f^{\prime}(x_{2^{q}(i-1)+2^{p+q},2^{p+q}}^{*})\end{array}\right)^{-1}\]
has already been calculated (see \ref{eq:matrizM}) and\[
N=\left(\begin{array}{c}
-x_{2^{q}(i-1)+2,2^{p+q}}^{*}+{\displaystyle \frac{1}{2^{p}}\sum_{i=1}^{2^{p+1}}}x_{2^{q}(i-1)+2,2^{p+q}}^{*}\\
-x_{2^{q}(i-1)+3,2^{p+q}}^{*}+{\displaystyle \frac{1}{2^{p}}\sum_{i=1}^{2^{p+1}}}x_{2^{q}(i-1)+3,2^{p+q}}^{*}\\
\vdots\\
-x_{2^{q}(i-1)+1,2^{p+q}}^{*}+{\displaystyle \frac{1}{2^{p}}\sum_{i=1}^{2^{p+1}}}x_{2^{q}(i-1)+1,2^{p+1}}^{*}\end{array}\right)\]
After operating in (\ref{eq:veinte}) it results in: \[
A_{k}=\frac{\alpha}{(-1)^{2^{p+q}}(1-[f^{2^{p+q}}(x_{1}^{*})]^{\prime})}\sum_{j=1}^{2^{p+q}}\left[f^{2^{p+q}-j+k-1}(x_{j+1,2^{p+q}}^{*})\right]^{\prime}\left(\left(x_{1+j,2^{p+q}}^{*}\right)+\frac{1}{p}S_{j}\right)\]
\[
k=1,\,...,\,2^{p+q}\]
with\[
S_{j}=\left\{ \begin{array}{lcl}
{\displaystyle \sum_{i=1}^{2^{p}}}x_{2^{q}(i-1)+2,2^{p+q}}^{*} & \quad & \mbox{if}\quad j=[2^{q}]\\
{\displaystyle \sum_{i=1}^{2^{p}}}x_{2^{q}(i-1)+3,2^{p+q}}^{*} & \quad & \mbox{if}\quad j=[2^{q}]+1\\
{\displaystyle \sum_{i=1}^{2^{p}}}x_{2^{q}(i-1)+4,2^{p+q}}^{*} & \quad & \mbox{if}\quad j=[2^{q}]+2\\
{\displaystyle \sum_{i=1}^{2^{p}}}x_{2^{q}(i-1)+5,2^{p+q}}^{*} & \quad & \mbox{if}\quad j=[2^{q}]+3\\
\qquad\qquad\vdots & \vdots & \qquad\quad\vdots\\
{\displaystyle \sum_{i=1}^{2^{p}}}x_{2^{q}(i-1)+2^{q}+1,2^{p+q}}^{*} & \quad & \mbox{if}\quad j=[2^{q}]+2^{q}-1\end{array}\right.\]
where $[2^{q}]$ represents a multiple of $2^{q}$, with the periodicity
conditions \[
\begin{array}{c}
A_{i+2^{p+q}}=A_{i}\\
x_{i+2^{p+q},2^{p+q}}^{*}=x_{i,2^{p+q}}^{*}\end{array}\]
due to the cyclic character of the $2^{p+q}$-periodic orbit.

The solution obtained is valid at order $O(\varepsilon^{2})$ while
$\varepsilon\ll\frac{1}{1-[f^{2^{p+q}}(x_{1}^{*})]^{\prime}}$. The
saddle-node orbit ($[f^{2^{p+q}}(x_{1}^{*})]^{\prime}\neq1$) has
been avoided by the condition $q\in\mathbb{N}$ ($0\not\in\mathbb{N}$),
and therefore there has been, at least, one period-doubling bifurcation.

\end{description}

\subsubsection*{Remarks:}

\begin{enumerate}
\item Notice that this theorem indicates that a CML, with $2^{p}$ oscillators,
has originally a $2^{p}$-period travelling wave. As the $2^{p}$-periodic
orbit of $f$ duplicates ($q$ increases) to a $2^{p+q}$-periodic
orbit so does the travelling wave of the CML.
\item The theorem 5 does not impose any restriction to the $2^{p}$-periodic
orbit of $f$. In the case of $f$ presenting $2^{p}$-periodic windows,
this $2^{p}$-periodic orbit could belong to a period-doubling cascade
in the canonical window, or originate from a $2^{p}$-periodic saddle-node
orbit. In the former case, $f$ undergoes a period-doubling cascade,
in the latter it is $f^{2^{p}}$ who goes through a period-doubling
cascade (this period-doubling cascade would be located inside a $2^{p}$-periodic
window). The conclusion is straightforward: the CML will not have
just one $2^{p}$-periodic wave, undergoing a period-doubling cascade;
there will be as many as $2^{p_{1}}$-periodic windows of $f$, with
$p_{1}\leq p$ ($p_{1}=0$ would be the canonical window).
\item Similar arguments can be done for synchronized state cascades, subject
to the condition that the duplicating orbit does not have prime period:
it can be a $p\cdot q$-periodic orbit in the canonical window, or
a $q$-periodic orbit in the $p$-periodic window, or a $p$-periodic
orbit in a $q$-periodic window, that afterwards will undergo period
bifurcation cascade.
\item There is a fact that could be not observed at a first reading of theorem
5: the points used to construct the perturbative solution both can
be stable and unstable.
\item Since it has been deduced that the CML undergoes a period doubling
cascade and that this cascade has its origin in the period doubling
cascade of the $f$, it is concluded that the CML inherits the dynamics
of $f$. 
\end{enumerate}

\section{Discussion and conclusions}

Several theorems have been proved that show the sufficiency conditions
of existence of synchronized states (periodic and chaotic) and travelling
waves in CML. Also it has been analytically determined the value that
describes the state of each oscillator at any moment. The results
of the theorems are as general as possible. This is due to two facts.
Firstly, the CML, with which we have worked, has a number of arbitrary
oscillators. Second, the function $f$, that rules the dynamics of
every oscillator, is also arbitrary with the condition that it undergoes
a period-doubling cascade. 

The results have the following consequences and link with other research: 

\begin{enumerate}
\item [i)]The emergence of the global properties from the local ones has
been proved. The global dynamics inherits the dynamics of every oscillator:
fixed points of the system come essencially from the fixed points
of the map (that governs the dynamics of every oscillator) compounded
with itself $m\cdot2^{k}$ times (being $m$ number of oscillators
of the CML). In particular, this result has been observed recently
in numerical computations \cite{Palaniyandi}.

Our results are an explicit analytical expresion of the results of
Lemaitré y Chaté \cite{Lemaitre}, who proved, in CML, the traslation
of the local properties to a spatiotemporal level.

\item [ii)]The dynamics of a CML has been studied, where the individual
dynamics of every oscillator is ruled by an arbitrary function $f$;
being few the analytical results on the matter, one normally only
works with quadratic functions or piece-wise linear functions \cite{Atmans1,Anteneodo,Liyorke}.
The one presented here is an interesting generalization that permits
the calculation of properties associated with the states and their
evolution.
\item [iii)]Two limitations that are present when numerical techniques
are applied have been overcome:

\begin{enumerate}
\item Spurious results, due to finite precision in computer simulations
\cite{Grebogi,Czhou2000,Czhou2002}, have been avoided.
\item Limits tending to infinity can be used for analytical solutions, both
with the number of oscillators in CML and the number of bifurcations
in the period doubling cascade.

On the one hand, the numerical simulation with a large number of oscillators
becomes unaffordable due to the computation time that would be necessary
as the number of oscillators grows. On the other hand, as it has been
indicated in the introduction, the study of the onset of the turbulence
in a fluid to be properly understood would need many oscillators,
the more the better.

As a direct consequence, of taking the limit in the period doubling
cascade, it is deduced the existence of waves of arbitrary period,
tending to infinity as the parameter bifurcation gets closer to the
Myrberg-Feigenbaum point. This is a response to the established question
of Gade and Amritkar in their work \cite{Pmgade} where they found
the wavelength-doubling bifurcation. Another question raised by Gade
and Amritkar in that same paper was: is there more than one value
such that, if the parameter value tends to it, then the period of
the travelling wave tends to infinity?. The response again in the
afirmative; in fact, there are infinite values that are the correspondent
Myrberg-Feigenbaum points of the windows inside the canonical window.
The position of these values is determined by the Saddle-Node Bifurcation
Cascades \cite{Sm1} and the relation between period doubling cascade
and Saddle-Node Bifurcation Cascades is also known \cite{Sm2}. 

\end{enumerate}
\end{enumerate}

\end{document}